\documentclass[journal=jacsat,manuscript=article,layout=twocolumn]{achemso}

\usepackage[version=3]{mhchem} 
\usepackage[utf8]{inputenc}
\usepackage[T1]{fontenc}
\usepackage{graphicx}
\usepackage{amsmath}
\usepackage{multirow}
\usepackage{xcolor}
\usepackage{ulem}
\usepackage{makecell}

\title{Probing Peptide Adsorption Kinetics and Regioselectivity via Multipolar Plasmonic Modes of Gold Resonators}

\author{Mathieu Nicolas}
\affiliation{Sorbonne Universit\'{e}, CNRS, Institut des NanoSciences de Paris, INSP,  F-75252 Paris, France}
\alsoaffiliation{Sorbonne Universit\'{e}, CNRS, Laboratoire de R\'{e}activit\'{e} de Surface, LRS,  F-75252 Paris, France}
\email{mathieu.nicolas@insp.jussieu.fr}
\author{Shuhui Yang}
\affiliation{Sorbonne Universit\'{e}, CNRS, Institut des NanoSciences de Paris, INSP,  F-75252 Paris, France}
\alsoaffiliation{Sorbonne Universit\'{e}, CNRS, Laboratoire de R\'{e}activit\'{e} de Surface, LRS,  F-75252 Paris, France}
\author{Christophe Méthivier}
\affiliation{Sorbonne Universit\'{e}, CNRS, Laboratoire de R\'{e}activit\'{e} de Surface, LRS,  F-75252 Paris, France}
\author{Souhir Boujday}
\affiliation{Sorbonne Universit\'{e}, CNRS, Laboratoire de R\'{e}activit\'{e} de Surface, LRS,  F-75252 Paris, France}
\author{Bruno Gallas}
\affiliation{Sorbonne Universit\'{e}, CNRS, Institut des NanoSciences de Paris, INSP,  F-75252 Paris, France}

\begin{document}

\begin{abstract}
Efficient peptide adsorption on metasurfaces is essential for advanced biosensing applications. In this study, we demonstrate how ellipsometric measurements coupled with numerical simulations allow for real-time tracking of temporin-SHa peptide adsorption on gold metasurfaces. By characterizing spectral shifts at 660 nm, 920 nm, and 1000 nm, we reveal a rapid saturation of surface coverage after 3.5 hours, with a significant preferential adsorption at the resonator ends. Our approach provides a novel methodology for monitoring peptide binding, which could be applied to a wide range of biosensor designs.
\end{abstract}
\section{Keywords} 
Circular Dichroism; Mueller Matrix; Peptide; Plasmonics; Magneto-electric coupling; Kinetic; Biosensor

\section{Introduction}
The control of the adsorption of biological molecules on surfaces is crucial in many fields. In particular, some peptides such as temporin possess antibacterial or antifouling properties that could prevent the formation of biofilms, and their immobilization on surfaces would find applications in the medical field \cite{Costa2011,Nicolas2022,Primo2024,Caselli2024}. However, peptides are small biomolecules, typically 1-2 kDa, and monitoring their uptake can be challenging.

Plasmonic surfaces have been proposed for the detection of small amounts of analytes.  These surfaces contain gold resonators that exhibit high sensitivity to the adsorption of small molecules due to the sensitivity of Localized Surface Plasmon Resonance (LSPR) to changes in the dielectric environment. Traditionally, detection is based on monitoring the wavelength shift of the LSPR absorption band associated with a dipolar mode, and this technique is now commercially available. 

Besides dipolar modes, multipolar modes of plasmonic resonators have also been proposed to increase the sensitivity \cite{Bastús2016,Butakov2016,Giessen2010,Hastings2014,Gu2012,Atwater2011,Yen2012}. The increased sensitivity of multipolar modes mostly results from a combination of highly confined near-fields, sharper resonances, or from the large gradients of the near-fields \cite{mushroom2015,nanostars2015}. To accommodate multipolar modes, the plasmonic nanostructures must have complex shapes so that the multipolar modes are spectrally separated from the dipolar ones. Due to the optical properties of gold, the dipolar modes of such structures must be shifted towards the (infra)red region of the spectrum so that the multipolar modes can be excited independently of the dipolar ones. As a result of the reduced ohmic losses of gold at longer wavelengths, the dipolar modes have an enhanced response. The interplay between these effects (sharper multipolar modes but stronger dipolar response), as well as polarization effects associated with the complex shape of the resonators, make it difficult to draw a straightforward conclusion about the best configuration for molecular sensitivity. 

In this work, we use surfaces containing resonators hosting multipolar resonances, U-shaped resonators, which allow to study the relative sensitivity of different LSPR modes in a reduced spectral range thanks to the possibility to excite independently the different modes using different incident polarizations. We determine the sensitivity to the different modes against changes of ambient. This sensitivity is compared to that observed upon the absorption a peptide on the surface of the resonators. We show that a sensitivity to a peptide surface coverage as low as 0.6 $\%$ can be achieved which is mode dependent. We use to a numerical analysis based on the Modal Fourier Method (also referred to as Rigorous Coupled Waves Analysis) to investigate how the different modes exhibit different sensitivities to the absorption of peptides onto the resonators. In particular, it is proposed that the temporin show a preferential absorption at the end of the arms rather than on their sides.

\section{Experiments}
\subsection{Sensitivity of the metasurface to ambient variations}

\begin{figure}[htbp]
\centering\includegraphics[width=8.5 cm]{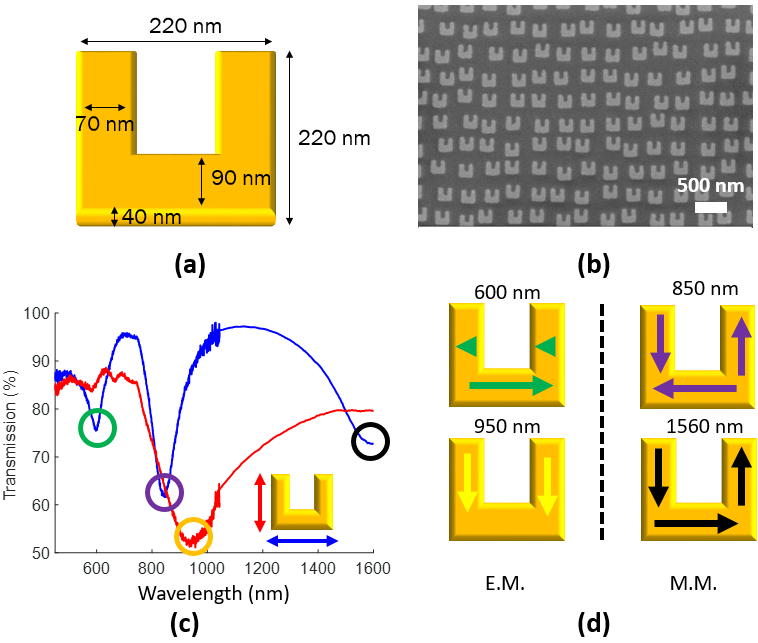}
\caption{(a) Schema of one U-shaped resonator and dimensions. (b) SEM image of the surface containing the resonators. (c) Transmission measurements in PBS of the metasurface. The color code of the incident polarization with respect to the U-shaped resonator orientation is given in the inset. The circles indicate the location of the different resonant modes observed. (d) Schema of the instantaneous current distributions in the U-shaped resonators for the resonant modes. The directions of the arrows give the current direction and their colors are associated with the colored circles of (c). The modes are separated in so-called ‘electric’ modes (E.M.) and ‘magnetic’ modes (M.M.).}
\label{fig:Samples}
\end{figure}

The metasurfaces consisted in glass substrates with U-shaped resonators made of gold deposited on it. The resonators were realized using a conventional e-beam lithography process. The resonators were deposited on a 4 nm Ti adhesion layer and had a nominal thickness of 40 nm. The dimensions of the resonators are depicted in Figure \ref{fig:Samples}(a). The resonators were localized on a pseudo-random lattice in order to reduce the signature of Rayleigh anomalies in the spectroscopic measurements. The pseudo-random lattice was obtained by changing randomly the position of the resonators initially located on a square lattice of 400 nm while ensuring that there would be no overlap of the resonators. Figure \ref{fig:Samples}(b) presents a scanning electron microscopy (SEM) image of the surface containing the resonators. 

The LSPR were first identified using conventional transmission measurement. To allow for an easier comparison with subsequent measurements, the sample was placed in Phosphate Buffer Saline (PBS) solution for the transmission measurements. Figure \ref{fig:Samples}(c) presents the result of the transmission measurements on the metasurface containing the U-shaped resonators. Two incident linear polarizations were used either parallel to the bottom arm of the resonators (blue line) or parallel to the lateral arms of the resonators (red line). Different dips associated with the excitation of resonant modes were observed indicated by circles. The current distributions at the resonances are depicted in Figure \ref{fig:Samples}(d). We find the usual electric and magnetic modes already described in the literature\cite{Proust2016, Amboli2023, Nicolas2023}. The mode around 600nm (green) will be called the electric quadrupolar mode, the one around 850nm (purple) the magneto-electric mode,  the one around 950nm (yellow) the electric dipole and the mode near 1560 nm (black) is the magnetic mode.

\begin{figure}[htbp]
\centering\includegraphics[width=8.5 cm]{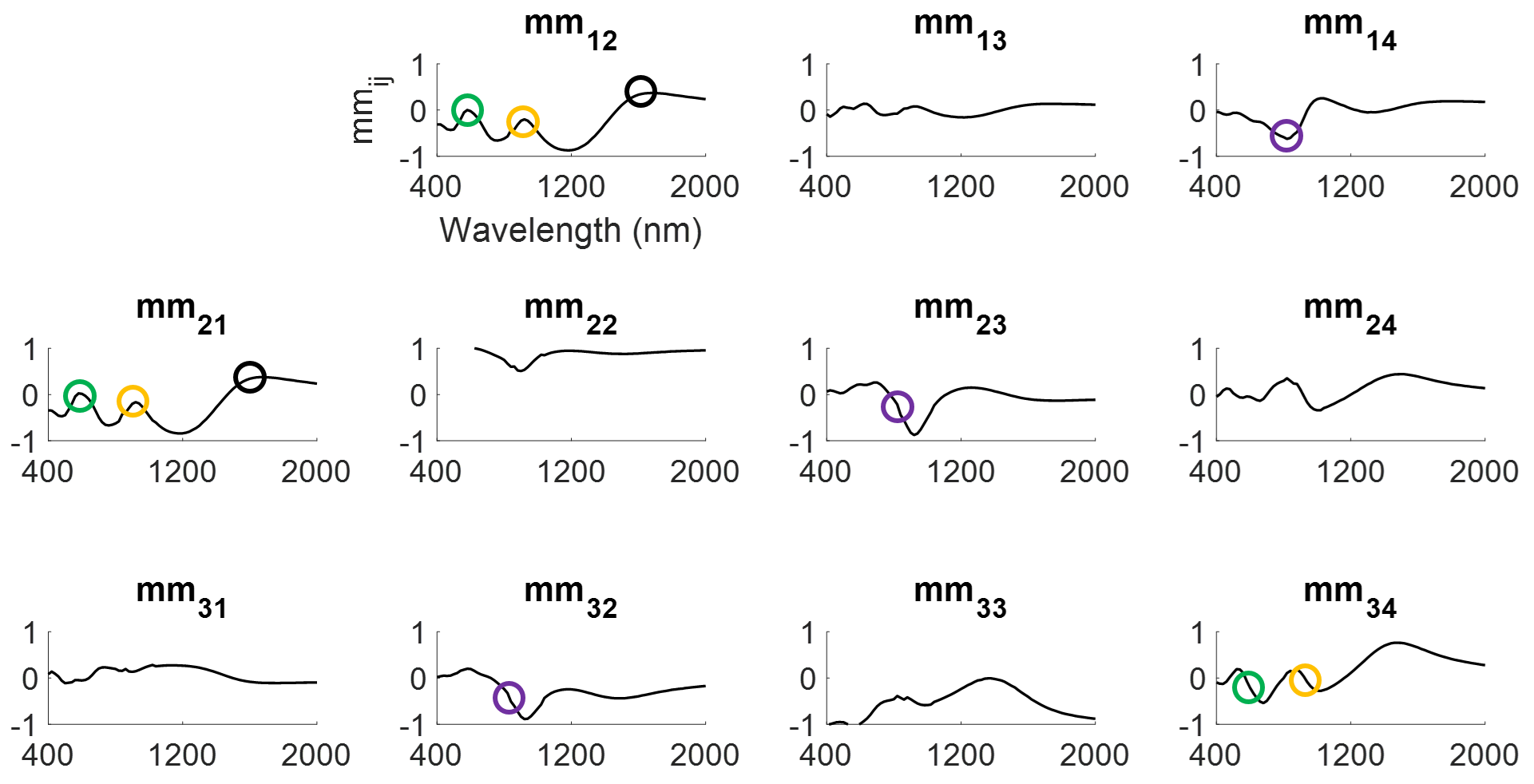}
\caption{Three first rows of the Mueller matrix of the metasurface with PBS as ambient. The circles indicate the position of the absorption bands observed in Figure \ref{fig:Samples}(c).}
\label{fig:MM}
\end{figure}

The polarimetric response of the metasurface upon grafting of peptides was measured in a liquid cell (see SI Figure S1), the resonators being in contact with the analyte. The liquid cell was placed on a V-VASE spectroscopic ellipsometer and the measurements were performed at a fixed incidence of 45$^o$ in the spectral range 400 nm - 2000 nm, with the incident beam from the glass substrate side, in order to determine the Mueller matrix of the metasurface. The Mueller matrix relates the incident polarization, projected on the Stokes vector, to the reflected one and contains all the polarimetric properties of the metasurface. The Stokes vector is defined as S=(I$_{tot}$, I$_{p}$-I$_{s}$, I$_{+45}$ – I$_{-45}$, I$_{RCP}$-I$_{LCP}$), where I$_{tot}$ is the total intensity, I$_{p}$ and I$_{s}$ are the intensities in p- and s-polarizations, I$_{+45}$ and I$_{-45}$ are the intensities for polarizations at +45° and -45° from p-plane and I$_{RCP}$ and I$_{LCP}$ are the intensities in right circular and left circular polarizations. As a consequence, the Mueller matrix is a 4x4 matrix. However, because our system has only a polarizer in the analyzer arm, only the three first rows were accessible and will be presented here. All the elements of the Mueller matrix were normalized to the first one: mm$_{11}$. The metasurface was aligned so that the plane of incidence contained the normal to the surface and the bottom arm of the resonators. For this alignment, the measurements were performed at an angle of incidence of 45° so as to maximize the off-diagonal elements of the Mueller matrix thanks to the excitation of the magneto-electric mode \cite{Proust2016}. Figure \ref{fig:MM} presents a typical measurement performed in PBS solution. The spectral locations of the absorption bands observed in the transmission measurements on the metasurface are indicated by circles with the same colours and positions as in Figure \ref{fig:Samples}. These absorption bands are associated with characteristic features in the Mueller matrix elements. It can also be noted that these features correspond to either extrema or large slopes of the spectral variations of the Mueller matrix elements. This is noticeable for mm$_{12}$ and mm$_{34}$ for instance were the signature of the electric modes are easily observable; for mm$_{23}$ and mm$_{14}$ were the signature of the magneto-electric mode is present. For example, the maximum in mm$_{14}$ correspond to a large slope in mm$_{23}$. This is reminiscent of the usual interpretation of the Mueller matrix elements in term of circular dichroism and circular birefringence  for mm$_{14}$ and mm$_{23}$, respectively.

\begin{figure}[htbp]
\centering\includegraphics[width=8.5 cm]{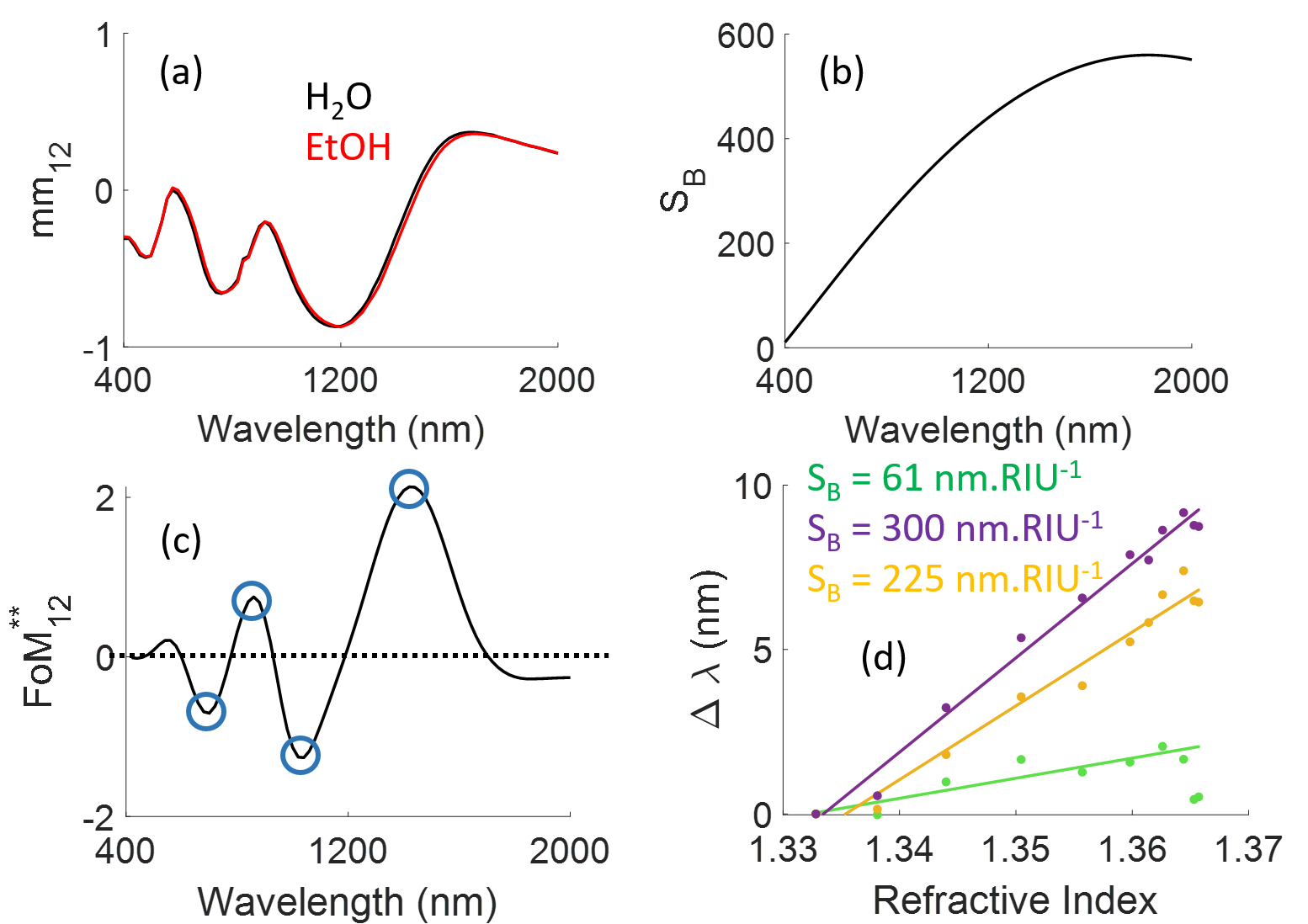}
\caption{(a) Evolution of the element mm$_{12}$ when changing the ambient from water to ethanol. (b) Sensitivity S$_{B}$ as a function of wavelength calculated from the spectral shift of mm$_{12}$ of (a). (c) Alternative Figure of Merit FoM$_{12}^{**}$ obtained from mm$_{12}$ as a function of wavelength. (d) Wavelength shift as a function of refractive index of the ambient for the quadrupolar mode, electric dipolar mode and magneto-electric modes.}
\label{fig:sensitivity}
\end{figure}

With the aim of monitoring the adsorption of peptides, the first calibration consisted in determining the evolution of the Mueller matrix elements upon modification of the dielectric constant of the ambient medium. This modification was obtained by testing PBS and ethanol as ambient, hence changing the refractive index from 1.336 to 1.36. Figure \ref{fig:sensitivity}(a) presents the modification of the Mueller matrix element mm$_{12}$. Similar variations were obtained for the other matrix elements (see SI Figure S2). It appears that the evolution corresponds to an overall shift $\delta \lambda$ of the spectral features of the Mueller matrix elements, although the shift was not uniform but increased linearly with wavelength (see SI Figure S3). From this spectral shift, the sensitivity S$_{B}$ was calculated as S$_{B}$($\lambda$) = $\Delta \lambda$ ($\lambda$)/ $\Delta$n($\lambda$) (see Figure \ref{fig:sensitivity}(b)), where $\Delta$n($\lambda$) was describing the spectral dependence of the wavelength difference between water (for PBS) and ethanol (see SI Figure 3 for all elements). To evaluate the most sensitive spectral regions, we have used an alternative Figure of Merit (FoM**) \cite{Sherry2005,Becker2010} derived from the usual FoM proposed for non-Lorentzian lineshapes and calculated as: FoM$_{ij}^{**}$ = S$_{B}$($\lambda$).$\delta$mm$_{ij}$/$\delta \lambda$, where $\delta$mm$_{ij}$/$\delta \lambda$ was calculated on the initial measurement performed in PBS. This FoM is directly normalized thanks to the normalization of the Mueller matrix elements and is presented in Figure \ref{fig:sensitivity}(c) for the element mm$_{12}$ (see SI Figure S5 for all elements). The extrema correspond to the spectral locations were the considered element is the most sensitive to variations in the refractive index of the ambient. These extrema are indicated by blue circles in Figure \ref{fig:sensitivity}(c). The sensitivity was then determined more accurately by changing more precisely the refractive index by mixing PBS with ethanol. The variations of $\Delta \lambda$ as a function of the refractive index variations $\Delta$n are presented in Figure \ref{fig:sensitivity}(d). The sensitivity S$_{B}$ is defined as the slope of these variations. The values were obtained by monitoring the variations of mm$_{34}$ at 600nm and 930 nm for the electric quadrupolar mode and electric dipolar mode, respectively and by monitoring the variations of mm$_{14}$ at 840 nm for the magneto-electric mode.

Depending on the mode considered, different values of S$_{B}$ were obtained ranging from 61 nm.RIU$^{-1}$ for the electric quadrupole mode at 600 nm to 300 nm.RIU$^{-1}$ for the magneto electric mode located at 840 nm. These values are in satisfactorily agreement with values found in the literature for other plasmonic systems \cite{Spackova2016}. The lower sensitivity of the electric quadrupole mode at 600 nm was expected since the wavelength shift $\delta \lambda$($\lambda$) increases with wavelength. However, more surprising is the lower sensitivity of the electric mode at 930 nm (S$_{B}$=225 nm.RIU$^{-1}$) as compared to the magneto-electric mode at 840 nm (S$_{B}$=300 nm.RIU$^{-1}$).
It must be noted that the larger value obtained for the magneto electric mode as compared to the electric mode, although it was monitored at a lower wavelength, reflects the larger value of the FOM$^{**}$ resulting from the larger spectral variations of the features associated with the magneto-electric mode as compared to the electric one. This result evidences the interest of being able to choose the element that can be monitored to have a better sensitivity, beyond the simple linear variation of $\delta \lambda$($\lambda$).

\subsection{Monitoring the adsorption kinetics of peptides}
\begin{figure}[htbp]
\centering\includegraphics[width=8.5 cm]{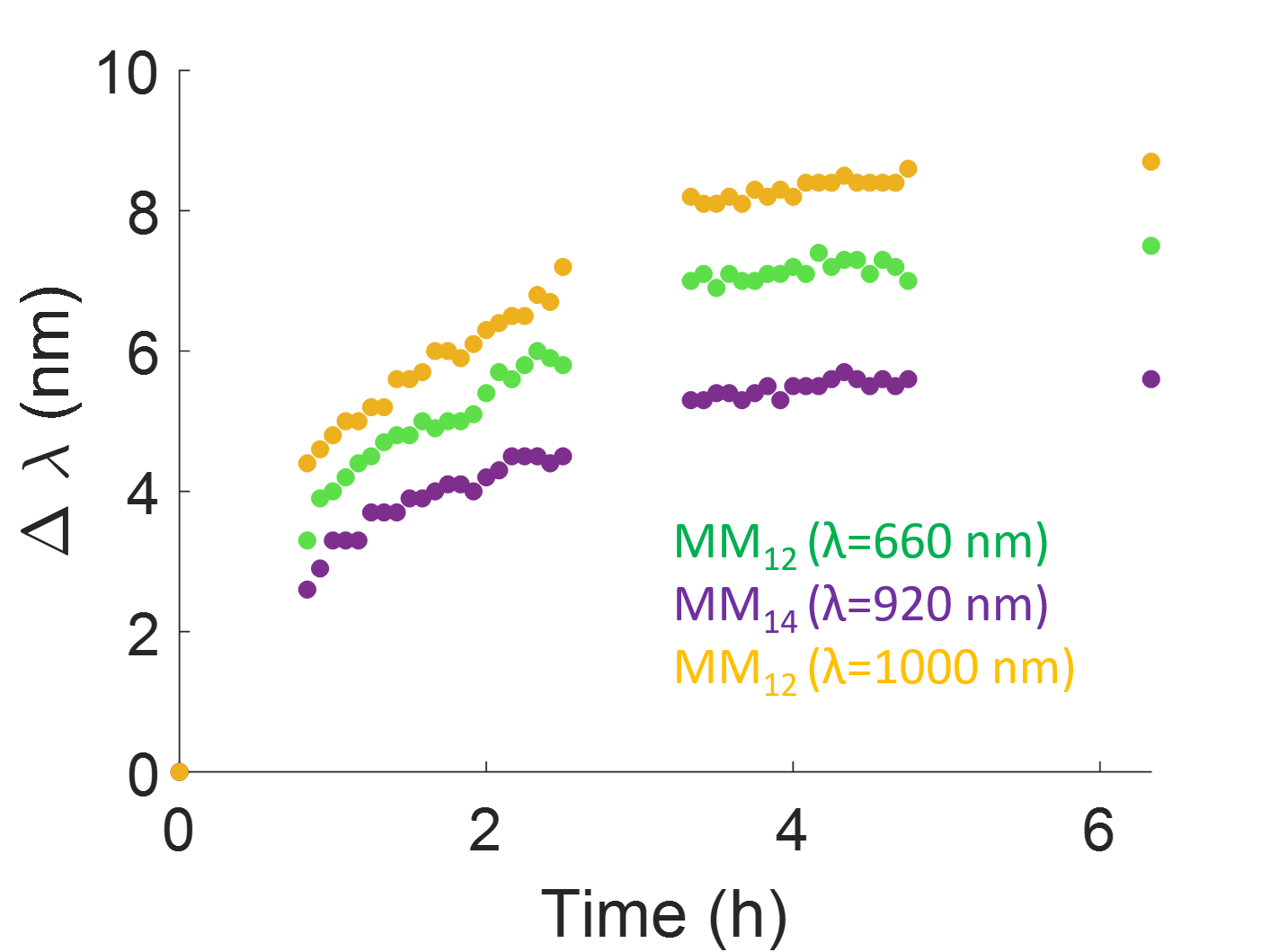}
\caption{Time dependence of the wavelength shift of the electric quadrupole (660 nm), magneto-electric mode (920nm) and electric dipole (1000 nm) during the adsorption of temporin on the metasurface.}
\label{fig:VarMM}
\end{figure}

The covalent immobilisation of temporin-SHa (FLSGIVGMLGKLF, 1.4 kDa) has been made on the characterized metasurfaces. In order to graft the peptides onto the gold surface, a direct approach was adopted, i.e. no cross-linker was used. With that aim, the peptides were first modified in order to obtain a thiolated temporin allowing a strong bonding to gold. To prove the adsorption of the peptide on
gold Polarization Modulated In-
frared Reflection-Absorption Spectroscopy (PM-IRRAS) has been realized (see SI Figure S6) \cite{Lombana2014,Vallée2009} . The peptides were diluted in PBS at 72.5 mg.L$^{-1}$ and introduced in a liquid cell to monitor the adsorption kinetics in static conditions. The volume of the liquid cell was of approximately 0.4 mL for a metasurface of 0.7 mm$^2$, which ensured that the kinetic was not limited by the amount of peptides available in the solution. Ellipsometric measurements were made every 5 minutes and the spectral variations in term of wavelength shift of the Mueller matrix elements mm$_{12}$ and mm$_{14}$ near 660 nm, 920 nm and 1000 nm are plotted in Figure \ref{fig:VarMM}. These values correspond to the extrema of the FOM$^{**}_{ij}$ and are associated with the resonant modes previously described. Full spectroscopic measurements were made at the beginning and after three and five hours, which explains the absence of points in these areas. The spectral shift of the different modes were of 7.5 nm for the quadrupolar mode ($mm_{12} (\lambda$ = 660nm)), 5.6 nm for the magneto-electric mode ($mm_{14} (\lambda$ = 920nm)) and of 8.7 nm for the electric dipole ($mm_{12} (\lambda$ = 1000nm)).

All Mueller matrix elements showed a fast increase at short time with a saturation after 3.5 h approximately (see Figure \ref{fig:VarMM}). This saturation corresponded to the saturation of the surface of the resonators by the temporin, meaning that after that period of time 3D aggregates of temporin start to form at the surface. This was also shown by Quartz Crystal Microbalance measurements (see SI Figure S7) \cite{Mazouzi2021,Boujday2008,Sauerbrey1959}, we can assume here that we are in a monolayer domain thanks to the static conditions. At saturation, the surface coverage was of approximately 40$\%$ as estimated from X-ray Photoelectron Spectroscopy (XPS) measurements (see SI Figure S8) which corresponds to approximately a mass of 3.3 10$^{4}$ zg of temporin per resonator. The variations in wavelength were as high as 8.7 nm when monitored on the element mm$_{12}$ at 1000 nm. The stability of the measurement measured over 10 h allowed for an accuracy of $\Delta \lambda$ = 0.1 nm (see SI Figure S9), yielding to a limit of detection 20 zg of temporin per resonator (surface coverage of 0.6$\%$). 

\subsection{Numerical analysis}
\begin{figure}[htbp]
\centering\includegraphics[width=8.5cm]{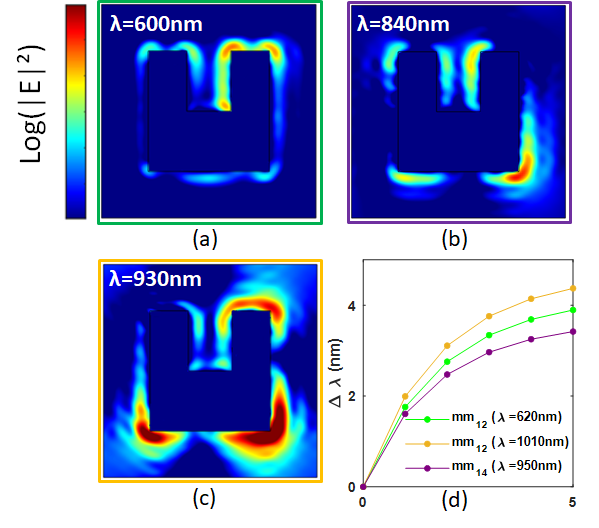}
\caption{Distribution of the electric field intensity around the resonators at (a) the electric quadrupole, (b) magneto-electric and (c) electric resonances investigated here. The intensity is presented of a logarithmic scale normalized to the value obtained without resonator. The color scale is the same for all maps. Spectral variations $\Delta \lambda$ calculated for characteristic Mueller matrix elements near the electric quadrupole (620 nm), magneto-electric mode (950 nm) et electric dipole (1010 nm) as a function of the refractive index value of the coating layer.}
\label{fig:nearfield}
\end{figure}

It may seem surprising that the spectral shift was approximately the same for the mode at 660 nm (quadrupolar mode) and the one at 1000 nm (electric mode) when measured on the same Mueller matrix element since the initial sensitivities were very different (Figure \ref{fig:sensitivity}(d)) : S$_{B}$=61 nm.RIU$^{-1}$ for the quadrupolar mode and S$_{B}$=225 nm.RIU$^{-1}$ for the electric mode. 
To investigate this point, numerical simulations were performed. 
They were based on the Modal Fourier Method using the code Simphotonic\_FMM.\cite{SimPhotonics} The geometrical parameters were slightly adjusted to yield a good agreement between the measured and calculated optical properties (see SI Figure S10). The periodicity of the structure was defined as a square unit cell with dimensions of 402 nm in both x and y directions. The structure composed multiple layers with different thickness, specifically, gold U-shaped resonator (37 nm) on a titanium adhesion layer (6 nm) and a glass substrate. The refractive indices of gold, titanium, and SiO$_2$ were obtained from well-established sources \cite{JC72,JC74}. The ambient medium was the PBS solution with a constant refractive index value of 1.330. This structure was illuminated from the glass substrate by a linearly polarized plane wave arriving at an angle of incidence 45$^o$. The optical response was calculated for an incident plane wave incoming in a plane containing the normal to the surface and the bottom arm of the resonator. First, the distribution of the electric field enhancement in a plane 20 nm above the glass surface was investigated. Figure \ref{fig:nearfield} presents the near field maps for the different modes.

\begin{figure}[h!]
\centering\includegraphics[width=8.5cm]{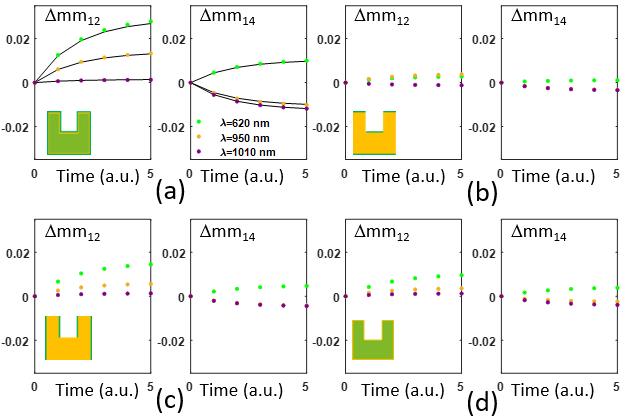}
\caption{Simulated variations of the elements mm$_{12}$ and mm$_{14}$ upon variation of the refractive index of a 1.2 nm thin layer covering the resonators. The coating layer was placed (a) on all faces of the resonators, (b) only at the end of top and bottom end of the arms, (c) only on the lateral sides of the arms, (d) only on the top surface of the resonators. The insets show a schematic of the resonator with in green the location of the thin film. For illustration purpose, the top layer does not fully cover the resonator in the insets. The black line in (a) is calculated as the sum of the values obtained in (b), (c) and (d). }
\label{fig:KinekSim}
\end{figure}

The incident field was polarized parallel to the bottom arms for investigating the modes at 600 nm (Figure\ref{fig:nearfield}(a)) and 840 nm (Figure\ref{fig:nearfield}(b)) and parallel to the lateral arms for the mode at 930 nm (Figure\ref{fig:nearfield}(c)).
It can be seen that the electric field enhancement is mostly located along the lateral sides of the arms at the electric quadrupole resonance and mostly at the ends of the arms at the electric dipole resonance. A mixed configuration is observed at the magneto-electric resonance. \\
To model the adsorption of temporin on the resonators, the resonators were uniformly coated with a thin layer (1.2 nm) with a refractive index varying between 1.33 and 1.49 in the numerical simulations. The maximum value of 1.49 corresponds to 40$\%$ of refractive index 1.7 (peptides) in water. We have arbitrarily chosen variation of the refractive index that would mimic the absorption kinetics observed in Figure \ref{fig:VarMM}. From the complex reflection coefficients, we have calculated the Mueller-Jones matrix at each wavelength. Figure \ref{fig:nearfield}(d) presents the spectral variation for the same Mueller matrix elements as presented in Figure \ref{fig:VarMM}. A general good agreement was observed between the simulated variations and the experimental ones. A similar observation could be made for all Mueller matrix elements and at different wavelengths (see SI Figure S11), where the direct variations of the Mueller matrix elements rather than the shift in wavelength are presented. The model assumed that the temporins were uniformly distributed over all the faces of the resonators so that the field enhancement at the quadrupolar mode (along the sides of the resonators) would probe the same density of species as the field enhancement at the electric and magneto-electric modes (near the end of the arms). The results upon variation of the location of the coating layer for different coverage configurations are presented for the elements mm$_{12}$ and mm$_{14}$ in Figure \ref{fig:KinekSim} . Figure \ref{fig:KinekSim}(a) corresponds to a configuration were the coating layer was all around the resonators. Figure \ref{fig:KinekSim}(b) correspond to a configuration where only the end faces of the arms were coated. The case where only the lateral sides of the resonators were coated yielded the variations is presented in Figure \ref{fig:KinekSim}(c). Finally, in the fourth configuration, only the top surface of the resonators was coated (Figure \ref{fig:KinekSim}(d)).

A closer inspection showed that the variations associated to the dipolar mode (yellow mode) were generally underestimated in the simulations with full coverage of the resonators (Figure \ref{fig:KinekSim}(a)) as compared to the measurements (Figure \ref{fig:VarMM} and SI Figure S11 for the values in Mueller units). In contrast a fairly good agreement was obtained for the electric quadrupole and magneto-electric modes. We can notice that the total variation in the Mueller matrix elements observed in Figure \ref{fig:KinekSim}(a) resulted directly from the sum of the variations associated with the complementary coverage configurations of Figures \ref{fig:KinekSim}(b-c). This could be observed for all the Mueller matrix elements (see SI Figure S11).

\begin{table}
\resizebox{\linewidth}{!}{

\centering
\begin{tabular}{|c|c|c|c|c|c|c|}
\hline
\multirow{3}{*}{Modes} & \multicolumn{4}{c|}{Relative contribution $\%$}\\ 
\cline{2-5}
& MM element & \makecell{End\\contribution} & \makecell{Lateral\\contribution} & \makecell{Top\\Contribution}  \\
\hline
\hline
Geometrical & - & 23 & 37 & 40 \\
\hline
 EQ & $mm_{12}$ & 8 & 55 & 35 \\
\hline
 ED & $mm_{12}$ & 29 & 42 & 27 \\
\hline
 ME & $mm_{14}$ & 21 & 39 & 40 \\
\hline
\end{tabular}
}
 \caption{Relative variations of selected Mueller matrix elements associated with the excited modes, extracted from Figure \ref{fig:KinekSim}. The relative contributions of the different faces to the total surface of the resonator is recalled in the first row.}
 \label{table:1}
\end{table}

Table \ref{table:1} presents a comparison between the geometrical surfaces of the end, lateral and top faces to the total available surface and the relative contribution of the modes to the variation of the Mueller matrix elements. The total did not add up to exactly 100$\%$ because the resonator environment was altered depending on the positioning of the peptide layer in the simulation. Interestingly, it can be seen in Table \ref{table:1} that the relative contribution of the lateral faces of the arms exceeded the relative area of those faces. This was particularly true for the electric quadrupole mode. The electric dipole mode showed also an enhanced sensitivity to the end faces of the arms while the quadrupolar mode showed a very small sensitivity to the end faces. Finally, the magneto-electric mode showed a sensitivity mostly proportional to the relative geometrical contributions. This result can be used to understand the relative variations of the calculated Mueller matrix variations at the electric dipole and electric quadrupole modes. The comparison between the nearfield distributions (Figures \ref{fig:nearfield}(a-c)) and the contribution of the different configurations of peptide coverage to the total variation of the Mueller matrix elements (Figures \ref{fig:KinekSim}(b-d)) supports the origin of our observations. Indeed, the presence of a coating layer at the ends of the resonators did not induce a significant variation in the Mueller matrix elements, particularly for the element mm$_{12}$ as seen in Figure \ref{fig:KinekSim}(b). In contrast, the element mm$_{12}$  seemed to be very sensitive to the presence of a coating along the lateral sides of the arms (Figure \ref{fig:KinekSim}(c)), particularly for the quadrupolar mode. The magneto-electric mode monitored in mm$_{14}$ seems rather insensitive to the location of the coating layer. All these observations are consistent with the distribution of the electric field in the nearfield presented in Figure \ref{fig:nearfield}. Indeed, the electric field enhancement was mostly located along the lateral sides of the arms for the quadrupolar mode (Figure \ref{fig:nearfield}(a)), at the end of the arms for the electric mode (Figure \ref{fig:nearfield}(c)) and a mixed situation was observed at the magneto-electric mode (Figure \ref{fig:nearfield}(b)). Finally, the variations observed in Figure \ref{fig:KinekSim}(d) confirm what is generally admitted, i.e. that the main contribution to the variations in plasmonic based sensors arise from the hotspots of the electric field which are mostly at the side faces of the arms and not at their surface (Figure \ref{fig:nearfield}).

These results suggest that the peptide coverage was different along the end faces of the arms as compared to the remaining surface of the resonators. In particular, our observations would be explained by a preferential adsorption at the end faces of the arms (or a larger refractive index in the simulations). The preferential adsorption could be favored by the presence of more defects, providing more adsorption sites, together with a more rounded shape, providing a better relaxation of steric constraints \cite{Wei2013}. It appears that measuring directly the spectroscopic variations of the full Mueller matrix provides a way of monitoring the location of the preferential adsorption of the peptides at the surface of the resonators.

\section{Conclusion}
However, we have demonstrated a novel method for monitoring the peptide adsorption kinetics on metasurfaces using ellipsometric measurements. Our results show that temporin-SHa adsorption follows a preferential binding at the ends of the resonator arms. Numerical simulations successfully reproduce the observed spectral shifts, providing insights into the field distribution and sensitivity of different resonant modes. The findings suggest that the spatial distribution of peptide adsorption significantly influences the sensor response. These findings open up new perspectives for the development of advanced biosensors, particularly for the detection of peptides, proteins and of biofilms in general, by offering precise and real-time monitoring of molecular interactions on metasurfaces.

\section*{Funding}
This work has been carried out thanks to the support of the ANR (n$^{\circ}$ ANR-23-CE42-0019).

\section*{Acknowledgment}
The authors acknowledge A. Miche from LRS for assistance with XPS measurements, M. Salmain from the Institut Parisien de Chimie Moléculaire (IPCM) for fruitful discussions on biological topics, and V. Humblot from Femto-FT for assistance with the adsorption of temporins on gold surfaces.

\section*{Disclosures}
The authors declare no conflict of interest in this work.

\end{document}